\documentclass[iop]{emulateapj}
\usepackage{epsfig, natbib} 
\usepackage{mathrsfs,amssymb}
\usepackage{graphicx,latexsym}
\newcommand{\hii}{{\sc Hii}}
\newcommand{\msol}{\mbox{${\rm M}_\odot$}}
\newcommand{\mup}{m_{\rm up}}
\newcommand{\mlo}{m_{\rm lo}}
\newcommand{\Mup}{M_{\rm up}}
\newcommand{\Mlo}{M_{\rm lo}}
\def\spose#1{\hbox to 0pt{#1\hss}}
\def\dt{\spose{\raise 1.0ex\hbox{\hskip2pt$\mathchar"201$}}}    

\slugcomment{*** Accepted to ApJ Letters Aug 10, 2011 ***}

\shorttitle{Origin of Salpeter Slope}
\shortauthors{Oey}

\begin{document}

\title{On the Origin of the Salpeter Slope for the Initial Mass Function}

\author{M. S. Oey}
\affiliation{Department of Astronomy, University of Michigan, 830 Dennison Building, 500 Church Street, Ann Arbor, MI, 48109-1042, USA}



\begin{abstract}

We suggest that the intrinsic, stellar initial mass function (IMF)
follows a power-law slope $\gamma=2$, inherited from 
hierarchical fragmentation of molecular clouds into clumps and clumps
into stars.  The well-known, logarithmic Salpeter slope $\Gamma=1.35$ in
clusters is then the aggregate slope for all the star-forming clumps
contributing to an individual cluster, and it is steeper than the
intrinsic slope within individual clumps
because the smallest star-forming clumps contributing to any given
cluster are unable to form the highest-mass stars.  Our Monte Carlo
simulations demonstrate that the Salpeter power-law index is the
limiting value obtained for the cluster IMF when the lower-mass limits
for allowed stellar masses and star-forming clumps are effectively
equal, $\mlo = \Mlo$.  This condition indeed is imposed for the
high-mass IMF tail by the turn-over at the characteristic value
$m_c\sim 1\ \msol$.  IMF slopes of $\Gamma\sim 2$ are obtained if
the stellar and clump upper-mass limits are also equal $\mup = \Mup
\sim 100\ \msol$,
and so our model explains the observed range of IMF slopes between
$\Gamma\sim 1$ to 2.  Flatter slopes of $\Gamma =1$ are expected when
$\Mlo > \mup$, which is a plausible condition in starbursts, where
such slopes are suggested to occur.  While this model is a simplistic
parameterization of the star-formation process, it seems likely to
capture the essential elements that generate the Salpeter tail of the
IMF for massive stars.  These principles also likely explain the 
IGIMF effect seen in low-density star-forming environments.

\end{abstract}

\section{Introduction} 

The mass distribution of stars at birth, known as the initial mass
function (IMF), is perhaps the most important fundamental
parameterization of the star formation process (e.g., Lada 2009; McKee
\& Ostriker 2007; Adams 2003).  When Salpeter (1955) first empirically
evaluated the IMF, he identified a simple power-law distribution for
stellar mass $m$, having exponent $\gamma = 2.35$:  
\begin{equation}
n(m)\ dm \propto m^{-\gamma}\ dm \quad .
\end{equation}
While it is now clear that the IMF has a
characteristic mass below which its form turns over, yielding
additional distinct regimes at lower masses (e.g., Kroupa 2001;
Chabrier 2003), Salpeter's original prescription nevertheless has
remained surprisingly robust for $m \gtrsim 1\ \msol$.  Since 
the stellar light from galaxies is dominated by these more luminous
stars, the ``Salpeter IMF'' has proven vital to the entire field of
galaxy evolution, in addition to star formation and stellar
populations. 

Yet, more than half a century after Salpeter's (1955) seminal work,
the physical factors yielding the value of $\gamma=2.35$ for the power-law
index remain elusive and poorly understood.  The origin of the IMF
continues to be a topic of intense discussion, and the reader is
referred to recent reviews by Lada (2009), Clarke (2009), Elmegreen
(2009), McKee \& Ostriker (2007), Bonnell, Larson \& Zinnecker (2007)
for comprehensive overviews on IMF theory.  

One obstacle to gaining deeper physical understanding is the fact that
the Salpeter slope emerges from a variety of simulations dominated by
different mechanisms.  In particular, both the core collapse (e.g.,
Krumholz et al. 2010) and competitive accretion scenarios (e.g.,
Bonnell et al. 2003) are able to reproduce the Salpeter 
slope.  Effects that are explored in modern simulations include
decaying vs driven turbulence, isothermal vs non-isothermal equations
of state, and inclusion or exclusion of magnetic fields, among other
factors; the IMF slope generally does not provide a strong
discriminant on this large parameter space (e.g., Clarke 2009; McKee \&
Ostriker 2007). 

Here, we suggest an explanation of the Salpeter slope that may
provide the simplest basic physical understanding, but which may
correspond to a framework within which the more specific physics of
actual star formation operates.

\section{From the cluster mass function to the IMF}

Most, if not all, stars form in clusters (e.g., McKee \&
Ostriker 2007; Lada \& Lada 2003), which in turn form
with an initial cluster mass function (ICMF).  Like the stellar IMF,
the cluster ICMF is also described well by a power-law distribution in
cluster mass $M$: 
\begin{equation}
N(M)\ dM \propto M^{-\beta}\ dM \quad .
\end{equation}
Again like the stellar IMF, the ICMF power law index is also found to
be robust and largely invariant, but slightly shallower, $\beta\sim 
2.0$, than the Salpeter slope, based on clusters covering a wide range of
scales, including super star clusters (e.g., Meurer et al. 1995),
massive and open clusters (e.g., Elmegreen \& Efremov 1997; 
Zhang \& Fall 1999; de Grijs et al. 2003), 
sparse OB groups (Oey et al. 2004), and the \hii\ region luminosity
function (e.g., Oey \& Clarke 1998; Kennicutt et al. 1989).  

The similarity in the slopes of the IMF and ICMF, $\gamma = 2.35$ vs
$\beta = 2.0$, respectively, has been previously noted (e.g.,
Elmegreen 2006).  Since star formation is a
hierarchical process, with smaller units fragmenting from larger ones,
we can examine the origin of the IMF as the relationship between the
IMF and ICMF.  The ICMF slope $\beta=2$ is a value that seems simpler
to understand.  As is often pointed out, this power-law index
corresponds to a uniform distribution of power between large and small
scales (e.g., Elmegreen 2006).  Furthermore, Zinnecker (1982)
showed that an IMF slope $\gamma =2$ results simply from Bondi-Hoyle
accretion $\dt{m}\propto m^2$ among the protostellar masses, with the mass 
function $n(m,t)$ evolving from some arbitrary initial distribution:
\begin{equation}
\frac{dn}{dt} + \frac{d}{dm}(\dt{m}n) = 0 \quad .
\end{equation}
This yields a simple power law IMF having $\gamma=2$ (see also Bonnell
et al. 2007; Hsu et al. 2010).  Thus $\gamma=2$ is a value that is intuitive, with
simple and reasonable physical bases from both the parent ICMF
equipartition arguments and build-up from simple accretion.  So then
why is the Salpeter slope of the IMF slightly steeper than $\gamma=2$?  

Elmegreen (2009) summarizes the three approaches to generating the
IMF:  1) Fragmentation, in which the core mass function and IMF is a
direct mapping from the parent cloud fragmentation; 2) Accretion, in
which the IMF is the product of protostar accretion processes not
necessarily linked to how clouds fragment; and 3) Interruption of 1)
or 2), in which the form of the IMF is also determined by factors
that preferentially limit fragmentation or accretion for high or
low-mass stars.  If we adopt our argument above that, simplistically, the IMF
should inherit $\gamma = 2$ from the ICMF according to both 1) and
2), then 3) is a likely candidate to explain the deviation of the
Salpeter slope:  there is likely some process that limits the formation
of high-mass stars in favor of low-mass ones.

In a fully hierarchical star formation scenario, the masses of the
parent clouds, and hence, cluster masses, are determined before the
masses of the star-forming clumps, and in turn, the constituent stars.
If the smallest clumps have masses on the order of stellar masses, then
the formation of high-mass stars will necessarily be limited in such clumps,
relative to the low-mass stars.  In recent years, it has been proposed
that the integrated galaxy IMF (IGIMF) has a steeper slope than the
IMF because the most massive stars may not form in
smallest clusters (Kroupa \& Weidner 2003; Weidner \& Kroupa 2005,
2006; see also Elmegreen 1999), causing the maximum stellar mass
$\mup$ to depend on the cluster mass.  This would cause the IGIMF to
favor lower-mass stars. 

We propose that the difference between the Salpeter slope $\gamma =
2.35$ and the ICMF slope $\beta=2$ is caused by a similar situation
governing the distribution of stellar masses within individual
clusters.  If, following a fully hierarchical scenario, stellar
subgroups are formed out of clumps with pre-determined masses, 
and if the clump mass distribution also generates a dependence of
$\mup$ on clump mass $M$, then the resulting IMF for the entire
cluster will be slightly steeper than the clump mass function.  This
scenario has essentially been proposed by Elmegreen (1997), who
presented this as random sampling from a fractal mass distribution.
Here, we present a parameterization that is much simpler than
Elmegreen's model, and we show that {\it the Salpeter slope simply
  results from the overlap between the stellar mass range and the
  clump mass range.}

\section{Relative mass ranges of clumps and stars}

We model the generation of stars in a cluster as the sum of
all stars forming from individual clumps in the parent molecular
cloud.  We assume that the hierarchical fragmentation of the cloud
into clumps follows the same $\beta=2$ power-law distribution found
for the ICMF, and further, that the fragmentation of clumps into stars
does the same, arguing that self-similar physical mechanisms govern
all these fragmentation processes on
smaller scales.  This breaks down around the characteristic value 
of $m_c\sim 1\ \msol$, suggested to be linked to the Jeans
mass (e.g., McKee \& Ostriker 2007; Clarke 2009).  Since we are
interested only in the upper IMF, however, scale-free cloud
fragmentation plausibly dominates in this regime.  But we note that
our model contrasts with those in which the IMF slope results from
physics that causes fragmentation into a steeper core mass
distribution (e.g., Padoan \& Nordlund 2002).

We construct Monte Carlo simulations drawing the clump mass function
from the $\beta=2$ power law, and then the stellar mass function
within each clump from the same distribution, within lower and upper
stellar mass limits $\mlo=$ and $\mup$.  The entire
cluster is then the aggregate of all the stars formed in all the
clumps, and the IMF is the composite for this aggregate.  
Our default sampling algorithm allows rejection of stars that cause the clump
mass limit to be exceeded, and continues sampling until the total
stellar mass is within $\mlo$ of the specified clump mass.
This is similar to, but a bit more strongly sorted, than the
``sorted sampling'' algorithm of Weidner \& Kroupa (2006; see also
Elmegreen 2006; Parker \& Goodwin 2007).  

\begin{figure}[h]
\includegraphics[width=7cm,angle=90]{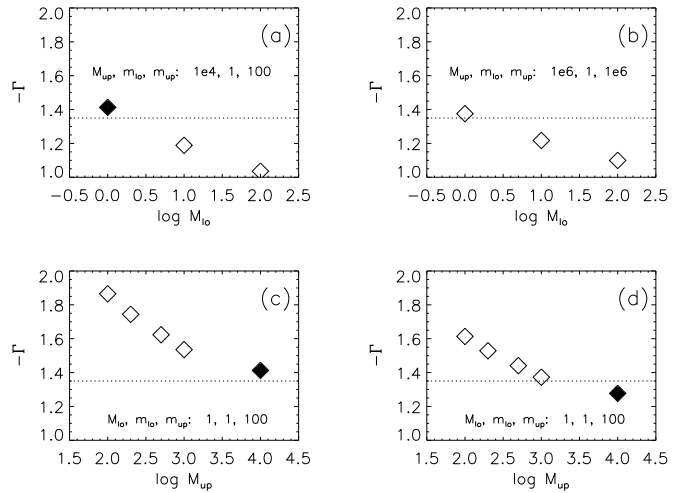}
\caption{Dependence of fitted logarithmic IMF slope $\Gamma$ on the
  allowed mass range for stars ($\mlo$ to $\mup$) relative to that
  for clumps ($\Mlo$ to $\Mup$).  Panels $a$ -- $c$ are based on
  sorted sampling, and panel $d$ is based on unsorted sampling.  To
  ease comparison, the black symbol shows the same model
  parameters, and the dotted line shows the Salpeter slope value.} 
\label{f_slopemodels}
\end{figure}

Figure~\ref{f_slopemodels} shows how the observed logarithmic IMF slope $\Gamma$,
determined from fits weighted by the inverse of the poisson errors,
depends on the mass ranges for the stars relative to the clumps.  
If the clump masses are all larger than the allowed stellar masses,
then the stellar masses are unconstrained, and the parent $\beta=2$
distribution is reproduced, translating to $\Gamma = 1$ on a
logarithmic plot.  This is apparent in Figure~\ref{f_slopemodels}$a$,
having a fixed allowed stellar mass range of 1 to 100 $\msol$, and
clump upper-mass limit $\Mup = 10^4\ \msol$.
Figure~\ref{f_slopemodels}$a$ shows $\Gamma$ as a function of the lower 
clump limit $\Mlo$, and as $\Mlo \rightarrow \mup$, $\Gamma\rightarrow
1$.  {\it But as allowed cloud masses decrease to values below $\mup$, the
IMF slope steepens, since formation of the most massive stars is
excluded.}  We see that as $\Mlo\rightarrow\mlo$, the value 
approaches $\Gamma = 1.4$, essentially the Salpeter value (black point).
Figure~\ref{f_slopemodels}$b$ confirms that this effect is driven by the allowed clump
lower-mass limit, rather than the upper limits on either the clumps or
stars.  This model allows both $\Mup$ and $\mup$ to be essentially
limitless, $10^6\ \msol$; the results are almost identical to
Figure~\ref{f_slopemodels}$a$ in approaching the same Salpeter slope
as $\Mlo\rightarrow\mlo$. 
This steepening effect between $\beta=2$ and the Salpeter value can
be seen in models by Elmegreen (1997) and Weidner \& Kroupa (2006).
Indeed, Elmegreen's model allowed
an effective equivalent $\Mlo\sim 1\ \msol$, and Weidner \& Kroupa
adopted an analogous $\Mlo = 5\ \msol$; both of these clump lower-mass
limits are well within the regime where the IMF steepening is expected,
according to our simulations.

IMF slopes steeper than the Salpeter value can be obtained if {\it
both} $\Mlo\ll \mup$ and $\Mup$ approaches stellar values.
Figure~\ref{f_slopemodels}$c$ assumes $\Mlo = 
\mlo$, and shows $\Gamma$ as a function of $\Mup$.  For $\Mup\gg\mup$,
$\Gamma\sim 1.4$, as found before for the assumed $\Mlo = \mlo$.  As
the clump upper-mass limit is decreased, $\Gamma$ steepens,
reaching $\Gamma = 1.9$ for $\Mup = \mup = 100\ \msol$.  The
results are essentially identical even if $\mup$ is limitless,
although the maximum stellar mass will be limited by $\Mup$ at low
$M$.  Weidner \& Kroupa (2006) also show that the degree to which
the aggregate IMF steepens depends slightly on the sampling algorithm
used.  A ``sorted sampling'' algorithm, as we use, allows continued sampling
of lower-mass stars after the highest-mass stars can no longer fit
within the alloted mass, and naturally induces slightly steeper slopes than
algorithms that stop sampling as soon as any drawn star causes the
clump mass limit to be reached.  Figure~\ref{f_slopemodels}$d$ is the
same as panel $c$, but with unsorted sampling, confirming that the
sorting scheme is not a strong effect.

Thus we see that the slope steepening beyond the input $\beta=2$
value is driven by allowing clump masses 
$\ll \mup$.  In these lowest-mass clumps, the
highest-mass stars cannot form, thus causing the aggregate cluster to
slightly favor lower-mass stars beyond the conventional IMF.  
From a physical standpoint, this 
simply means that clumps too small to produce the
highest-mass stars can still produce low-mass stars.  Furthermore,
the Salpeter IMF results when $\Mlo \sim \mlo$ while randomly
sampling both stars and parent clumps from the same $\beta=2$
power-law distribution.  Thus ironically, the critical parameter for
steepening the IMF and the IGIMF is not $\Mup$, but rather $\Mlo$. 

Elmegreen (1997, 1999) developed a model for IMF generation based on
random sampling of stellar masses from hierarchical fractal structure.  Our 
model distills the effects seen in Elmegreen's model to the most
simplististic level, and it more directly captures the dominant effect
seen in that work, revealing $\Mlo$ as the critical parameter.

\section{Discussion}

The essential condition that drives the steepening of the IMF from 
$\gamma = 2$ is that $\Mlo \ll \mup$.  This need not be
interpreted literally, but the smallest clumps must be incapable of
producing the highest-mass stars; our model is independent
of star-formation efficiency, provided that it is essentially
constant.  To attain the Salpeter value, in 
particular, requires the equivalent of $\Mlo \sim \mlo$.  Because the
full IMF flattens strongly near a characteristic value  $m_c\sim 1\
\msol$, this limits the power-law form of the high-mass IMF to apply
only above this characteristic mass, thus $\mlo \sim m_c$, recalling
that here we consider only the upper, Salpeter IMF.  Furthermore,
because stellar objects continue to form with masses 1 -- 2 orders of
magnitude lower, clump fragmentation clearly continues to masses
below $m_c$ as well, and so we know that the relevant lowest-mass clumps
having $\mlo\sim m_c$ do form stars.  Thus, for the
power-law upper IMF, $\Mlo\leq \mlo$, which is exactly the
condition needed to induce steepening of the aggregate cluster IMF to
the Salpeter value in the power-law regime above $m_c$.  In other
words, $m_c$ acts as an effective lower cutoff to the IMF power-law
distribution for both clump and stellar masses, rendering $\Mlo
=\mlo$.  Indeed, C$^{18}$O observations of M17-SW show that molecular
clump masses follow a power law of $\Gamma=-0.72\pm 0.15$ down to at least  
1 $\msol$ (Stutzki \& G\"usten 1990), while the observed IMF in the
closely associated region M17-SWex shows $\Gamma=1.3\pm 0.2$ (Povich \&
Whitney 2010).

If clumps fragment according to a $\beta=2$ power law, then the range
of observed IMF slopes should generally not fall below this value.
Similarly, steeper IMF slopes reach $\Gamma \sim 2$ 
for $\Mup = \mup= 100\ \msol$, when
$\Mlo\sim \mlo$.  A number of authors have shown compilations of
measured cluster slopes, and $1\leq\Gamma\leq 2$ does appear to be
a fairly well-defined range of allowed slopes for the
high-mass IMF (e.g., Elmegreen 1999, Fig.~1; Bastian et al. 2010,
Fig.~2):  what appears to be a large scatter about $\Gamma=1.35$
(Elmegreen 1999) can also be interpreted as variation within the range
allowed by our model.

The basic premise of our model is that clumps fragment from clouds
according to the $\beta=2$ power law, and in particular, that stars
formed within each individual clump are still formed with a $\gamma=2$
power law.  Thus if newly-formed stars can be identified with their
natal sub-groups in the youngest star-forming regions, the IMF should 
appear to be closer to $\Gamma=1$ in such groups.  It would be
interesting to evaluate the IMF slopes for sub-groups within massive
star-forming regions, and to compare these to the aggregate cluster IMF.
Also, observations focusing on specific subregions within extremely
young clusters might be more likely to obtain IMF slopes closer to
$\Gamma=1$, while more complete measurements for entire clusters
might obtain the Salpeter value or higher.  These are difficult tests,
since dynamical evolution will quickly mix stars born from different
clumps.  

It is often suggested (e.g., Elmegreen 2004, 2009) that the IMF 
is slightly flatter in starbursts and rich, young, super star clusters.  
In our model, recovering an IMF slope of $\gamma=2$ occurs
naturally if the clump mass $\Mlo \geq \mup$.  The Arches cluster near
the Galactic Center is the best-known example of a flat IMF, with
the most recent measurements of the high-mass IMF of $\Gamma=0.91\pm
0.08$ (Kim et al. 2006) and $\Gamma=1.1\pm 0.2$ (Espinoza et
al. 2009); these values are consistent with our model.  Regions of
extreme star formation are expected to have higher thermal Jeans
masses owing to radiative feedback from massive stars (e.g., Larson
2005; Murray 2009) 
therefore raising the minimum clump mass $\Mlo$.  Our slope flattening
would only occur if the clump mass range is increased without a
commensurate increase in the allowed stellar mass range; and indeed,
$\mup$ appears to be remarkably constant and independent of star
formation environment (e.g., Oey \& Clarke 2005; Weidner \& Kroupa
2004). 

If the smallest clusters form out of single clumps, with no
further fragmentation, then the cluster IMF is the same as the clump
IMF and there is no steepening of the aggregate slope.  Hence, such 
single-clump clusters should show the parent $\gamma=2$ stellar IMF.
Further, if the smallest clusters do approach limiting masses that are
on the order of the largest clump masses, this should cause an
analogous steepening of the cluster mass function under these 
circumstances.  As described above, the ICMF generally has a slope
$\beta=2$, but it would be interesting to evaluate any trends.  
We also note that our model does not necessarily imply an IGIMF steepening
itself:  smaller clusters can still fragment into clumps yielding the
full stellar mass range for all, including the smallest, clumps.
So if the smallest clusters are still capable of forming stars up to
masses $\mup$, as suggested by the results of Lamb et al. (2010), then
no IGIMF steepening will occur.  Thus again, the lower clump mass
range relative to the upper stellar mass range determines the steepening
for both the cluster stellar IMF and the IGIMF.  {\it We do however suggest
that the IGIMF effect seen in low-density environments} (e.g., Lee et
al. 2009; Hoversten \& Glazebrook 2008) {\it is real and occurs because the
smallest clusters have masses below $\mup$.}

We stress that our model is a simplistic parameterization of
hierarchical star formation, but it may capture essential
elements that govern the generation of the Salpeter tail of the IMF.
In particular, our model is based on a scenario that is fully
hierarchical, with clouds fragmenting into clumps and clumps into
stars, with each fragmentation process based on a $\beta=\gamma=2$
power-law mass distribution.  ``Fragmentation'' need not be literal,
but simply means that the ultimate mass apportionment from the parent
to descendant units follows this power-law distribution, whether by
competitive accretion or by well-defined fragmentation.  The
steepening of the IMF to the Salpeter value is then induced by the
inability of the lowest-mass clumps to form the highest-mass stars.
Maschberger et al. (2010) have carried out a star-formation simulation
of a large volume, which follows the hierarchical fragmentation of ISM
into clusters and down to core-like sink particles.  Their results are
consistent with our scenario, showing that clusters are mergers of
products from multiple clumps, and that the aggregate IMF in such
mergers is slightly steeper than in the individual merging units.

\bigskip\bigskip\bigskip
\section {Conclusion}

We propose that the empirical, logarithmic $\Gamma=1.35$ Salpeter
power-law slope of the IMF for high-mass stars originates from a
universal, linear $\gamma=2$ IMF slope for 
stellar sub-groups formed from clumps within individual clusters.  Our
Monte Carlo simulations show that $\Gamma=1.4$ is the limiting value
of the IMF slope when the lower-mass limit for both clumps and stars
is equal, $\Mlo = \mlo$; this condition holds for the high-mass tail
of the observed IMF, since it turns over below the characteristic mass
$m_c\sim 1\ \msol$, effectively setting the lower-mass limit to both
$\mlo$ and $\Mlo$.  The steepening to the Salpeter value occurs
because the highest-mass stars cannot form in the lowest-mass clumps.
Thus, a critical factor to examine in star formation theory is the
effective mass range for clumps relative to the output stellar
masses.  Our model is analogous to the mechanism for steepening 
the IGIMF proposed by Kroupa \& Weidner (2003), but a universal
Salpeter slope does not necessarily imply that the IGIMF should
steepen to even higher values, since the smallest clusters still may
be capable of forming the highest-mass stars.  However, we stress that
the same principles involving the relative mass ranges of the parent
and descendant units also applies to the IGIMF and may be responsible for
the apparent steepening in regimes dominated by the smallest-scale
star formation (e.g., Lee et al. 2009).

Our model is based on a fully hierarchical scenario in which the
physics of fragmentation from large scales to individual stars takes
place in a self-similar manner, all with a --2 power-law mass
distribution.  Thus, this does not require literal fragmentation, but
simply that the final mass distribution among both the parent and
descendant units is distributed accordingly.  With this simple
condition, we show that the aggregate high-mass IMF slope is limited
between values of $\Gamma=1$ and 2, 
which is consistent with observations.  Flatter values near $\Gamma=1$
occur when $\Mlo > \mup$, which may explain the suggested IMF
flattening for regions of extremely intense, massive star formation.
While our model is only a rough parameterization of the star formation
process, we suggest that it may capture the fundamental effects that
generate the Salpeter slope of IMF for high-mass stars.

\acknowledgments

Thanks to Cathie Clarke, Joel Lamb, and Fred Adams
for useful discussions and comments, and to the referee, Michael Reid.
I gratefully acknowledge the Institute of Astronomy, Cambridge for
visitor support, and funding by NSF grant AST-0907758. 


\end{document}